\documentclass[conference]{IEEEtran}

\usepackage[english]{babel}

\usepackage[letterpaper,top=2cm,bottom=2cm,left=3cm,right=3cm,marginparwidth=1.75cm]{geometry}

\usepackage{amsmath}
\usepackage{balance}
\usepackage{graphicx}
\usepackage[colorlinks=true, allcolors=blue]{hyperref}

\begin{document}
\title{A Deep Learning Approach Towards Student Performance Prediction in Online Courses: Challenges Based on a Global Perspective}

\author{
\IEEEauthorblockN{Abdallah Moubayed\IEEEauthorrefmark{1}\IEEEauthorrefmark{2}\IEEEauthorrefmark{4}, MohammadNoor Injadat\IEEEauthorrefmark{3}, Nouh Alhindawi\IEEEauthorrefmark{1}\IEEEauthorrefmark{5}, Ghassan Samara\IEEEauthorrefmark{6},\\ 
Sara Abuasal\IEEEauthorrefmark{7}, and Raed Alazaidah\IEEEauthorrefmark{3}
} 
\IEEEauthorblockA{\IEEEauthorrefmark{1}School of Computing and Augmented Intelligence, Arizona State University, Gilbert, Arizona, USA;\\ e-mails: \{amoubaye, nalhinda\}@asu.edu}
\IEEEauthorblockA{\IEEEauthorrefmark{2}Electrical and Computer Engineering Department, Lebanese American University, Byblos, Lebanon;\\ e-mail: abdallah.moubayed@lau.edu.lb}
\IEEEauthorblockA{\IEEEauthorrefmark{3}Data Science and Artificial Intelligence Department, Faculty of Information Technology, Zarqa University, Zarqa, Jordan;\\ e-mails: 
\{minjadat, razaidah\}@zu.edu.jo}
\IEEEauthorblockA{\IEEEauthorrefmark{4}Electrical and Computer Engineering Department, Western University, London, Ontario, Canada;\\ e-mail: amoubaye@uwo.ca}
\IEEEauthorblockA{\IEEEauthorrefmark{5}Faculty of Science and Information Technology, Jadara University, Irbid, Jordan;\\ e-mail: hindawi@jadara.edu.jo}
\IEEEauthorblockA{\IEEEauthorrefmark{6}Computer Science Department, Faculty of Information Technology, Zarqa University, Zarqa, Jordan;\\ e-mail: 
gsamara@zu.edu.jo}
\IEEEauthorblockA{\IEEEauthorrefmark{7}Software Engineering Department, Faculty of Information Technology, Zarqa University, Zarqa, Jordan;\\ e-mail: 
sabuasal@zu.edu.jo}
}


%Telephone: (800) 555--1212, Fax: (888) 555--1212}


\maketitle

\begin{abstract}
Analyzing and evaluating students’ progress in any learning environment is stressful and time consuming if done using traditional analysis methods. This is further exasperated by the increasing number of students due to the shift of focus toward integrating the Internet technologies in education and the focus of academic institutions on moving toward e-Learning, blended, or online learning models. As a result, the topic of student performance prediction has become a vibrant research area in recent years. To address this, machine learning and data mining techniques have emerged as a viable solution. To that end, this work proposes the use of deep learning techniques (CNN and RNN-LSTM) to predict the students' performance at the midpoint stage of the online course delivery using three distinct datasets collected from three different regions of the world. Experimental results show that deep learning models have promising performance as they outperform other optimized traditional ML models in two of the three considered datasets while also having comparable performance for the third dataset.
\end{abstract}

\begin{IEEEkeywords}
	Deep Learning, e-Learning, Online Courses, Student Performance Prediction 
\end{IEEEkeywords}

% \fi
\IEEEpeerreviewmaketitle

\section{Introduction}
\indent\indent With the continued growth of Internet users, the demand for new learning paradigms that rely on e-Learning environments is increasing rapidly \cite{injadat2020multi,injadat2020systematic}. Different researches have been conducted in the e-Learning environment in order to improve the beneficial advantages of e-Learning courses’ delivery \cite{musleh2020learning,nouh2021,nouh2022}. The ongoing researches in this area are trying to address certain objectives that enhance the e-Learning environment from a certain point of view. One of those important aspects is offering personalized learning \cite{kausar2018integration}. The researchers focus on personalized learning because of the importance of this aspect in improving the e-Learning contents and delivery methods to satisfy the learner’s needs \cite{murtaza2022ai}. Another important aspect is the adaptive learning, which concentrates on the learning style of the learner \cite{kolekar2019rule}. Both aspects will encourage the e-Learning environment’s parties to integrate the required features and technologies in the e-Learning environment in order to satisfy the individual learner’s needs \cite{zhang2022identifying}. Consequently, this requires research in learning analytics to analyze the learners’ needs. Thus, conducting traditional analytical research with the large number of learners nowadays is a complex and time consuming process. \\
\indent As a solution, this paper proposes a predictive model to predict the learner’s final grade in the course at earlier stages during the course. The predictive model implements deep learning (DL) models to classify the students and predicts their final grades at the midway point of the online course. More specifically, a convolutional neural network (CNN) and a recurrent neural network with long short term memory (RNN-LSTM) models are proposed as they have the potential to accurately predict student performance given their promising results in other applications such as pancreas and breast tumor detection \cite{zavalsiz2023comparative,narayanan2022hybrid}. As a result, this will help the instructors at earlier stages of the course delivery to take care of the students who may need help. \\
\indent To evaluate the performance of the proposed deep learning model, three distinct datasets from three different universities located in three different regions in the world are used. More specifically, the first dataset is for a first year engineering course at a European University. The second dataset is for a third year engineering course at a North American University. Finally, the third dataset is for a first year undergraduate course in Information Technology (IT) at a Middle Eastern university. This provides us with a more global model that is applicable to different courses and student demographics.\\
\indent The rest of the paper is organized as follows: Section \ref{related} describes the previous researches and related work. Section \ref{proposed_frame} represents the proposed approach. Section \ref{data_desc} describes the datasets used. Section \ref{performace} evaluates the performance of the proposed framework in comparison with other works from the literature. Finally, Section \ref{conc} concludes the paper and discusses the future work.
\section{Related Works}\label{related}
\begin{figure}[!ht]
	\centering
	\includegraphics[scale=.45]{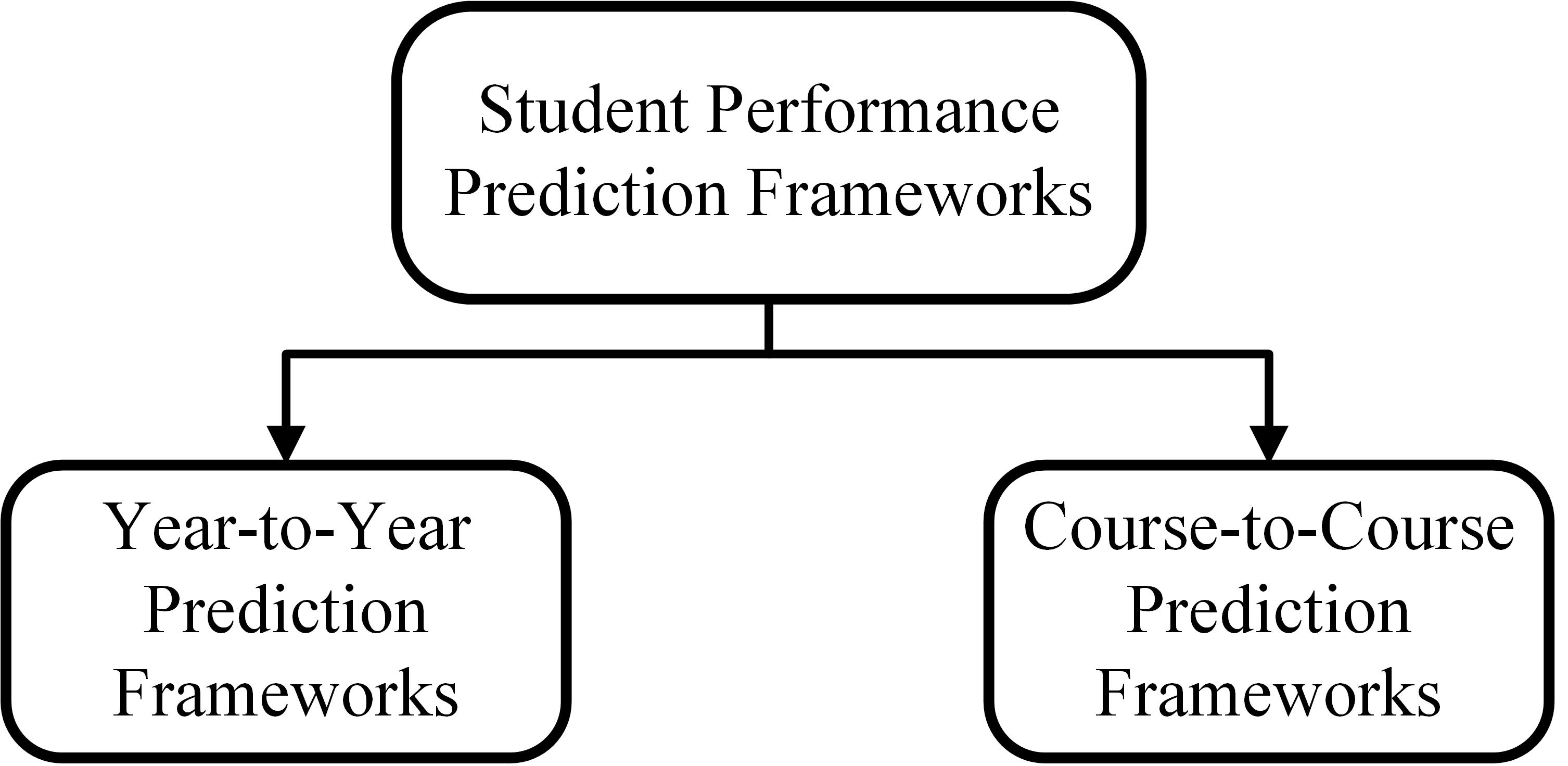}
	\caption{Student Performance Prediction Frameworks' Categorization}
	\label{related_categorization}
\end{figure}

\indent Due to the fact that this research area has been gaining significant attention and growing in popularity, several research works have investigated the use of machine learning and data mining techniques in educational settings. As shown in Figure \ref{related_categorization}, these works can be categorized as either being year-to-year prediction frameworks (such as \cite{ahmed2014data,khan2015final,saxena2015educational,jain2019efficient}) or course-to-course prediction frameworks (such as \cite{rana2016evaluation,wang2017learning}). Year-to-year prediction frameworks aim to predict the performance of the students in the courses/classes of a year based on their performance in the previous year. Alternatively, course-to-course prediction frameworks aim to predict the performance of students in a course based on their performance in previous similar courses.

\indent For example, the authors in \cite{ahmed2014data} proposed the use of classification models to predict the final grades of students. This was done by analyzing data from a course program across 6 years.\\
\indent Similarly, the authors of \cite{khan2015final} proposed using a decision tree algorithm, namely the J48 (also known as C4.5) algorithm, to predict student performance. Their results emphasized the potential of such algorithms for student performance prediction as they had high accuracy and speed.\\
\indent The authors of \cite{saxena2015educational} also proposed using the J48 algorithm by comparing its performance with that of the k-means algorithm. The results reiterated the superiority of the the J48 algorithm in predicting student performance.\\
\indent In contrast, the authors of \cite{rana2016evaluation} investigated the performance of different clustering algorithms such as k-means and hierarchical clustering in accurately predicting student performance. The experimental results showed that k-means algorithm outperformed other algorithms as it had the better performance and the faster building time.\\
\indent The authors in \cite{wang2017learning} proposed the use of deep neural network model for an e-learning recommendation framework. Experimental results illustrated that the proposed model improved the learning experience of the students.\\
\indent In a similar fashion, the authors of \cite{fok2018prediction} also proposed using deep learning to predict student performance. Their distinctive characteristic was that they used both academic and non-academic subjects. Experimental results showed that deep learning models are a viable potential solution for such problems as it can offer high accuracy.\\ 
\indent On the other hand, the authors in \cite{jain2019efficient} compared the performance of four tree-based models in accurately predicting students' performance. Using a Portuguese high school dataset, their experimental results showed that tree-based models achieved high accuracy and fast execution time.\\
\indent Finally, the authors in \cite{moubayed2020student,moubayed2018relationship} used unsupervised learning models to cluster users in various engagement levels. Based on these levels, they used Apriori association rules to related academic performance with student engagement. Their experimental results showed that there exists a positive correlation between students’ engagement level and their academic performance in an e-learning environment.\\
\indent Despite the promising results presented in previous related works, their main limitation lies in the fact that they rely on courses/classes that have already been completed to predict the performance in subsequent courses/classes. As mentioned earlier, this is done either on a course-to-course basis or a year-to-year basis. However, very few previous works aimed at predicting the student performance in a course during its delivery. Hence, this work aims at filling this gap by trying to accurately predict the performance of the students at the midpoint stage of the course delivery to ensure satisfactory completion of the course itself rather than in subsequent courses. 
\section{Proposed Framework} \label{proposed_frame}
\begin{figure}[!ht]
	\centering
	\includegraphics[scale=.4]{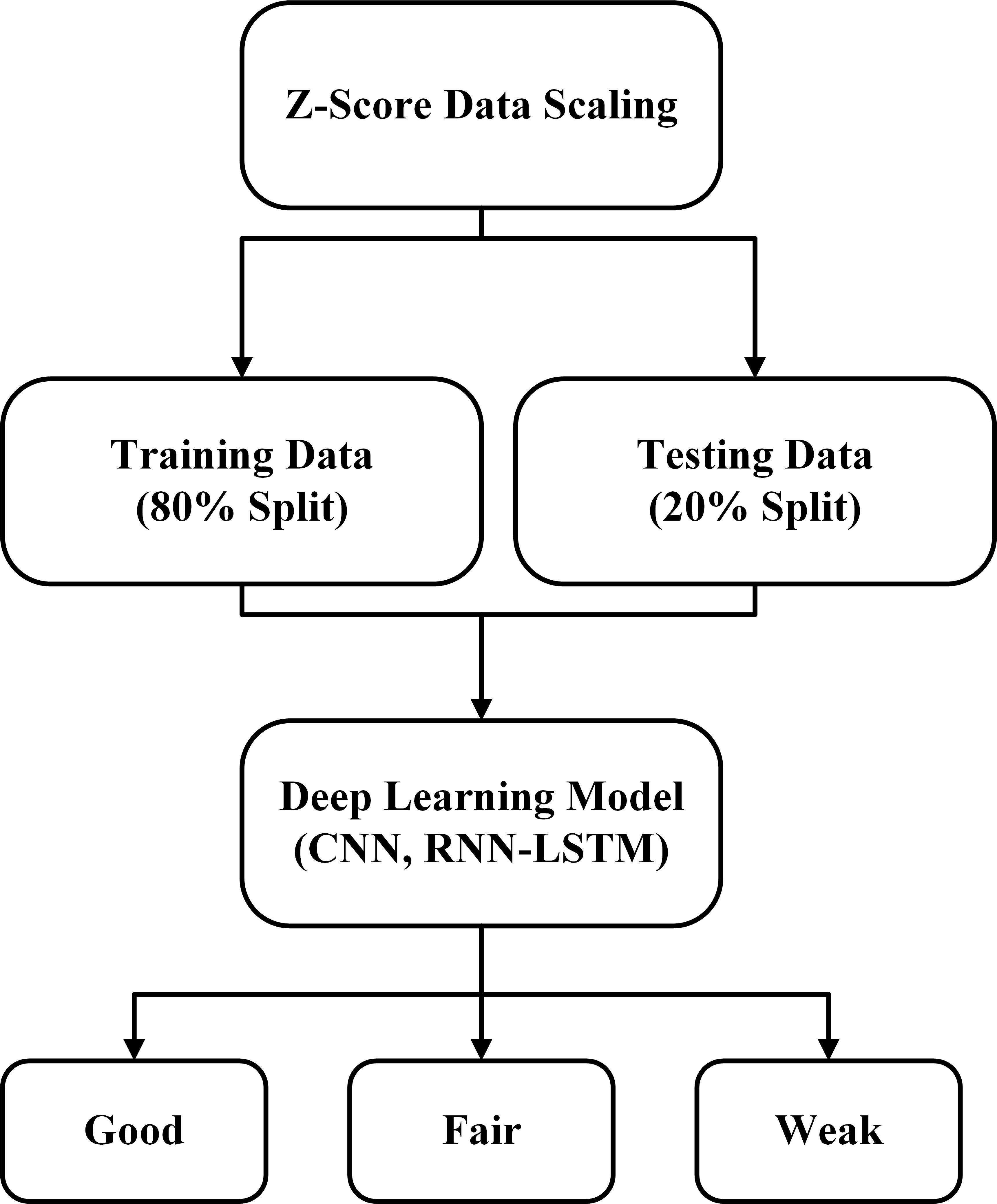}
	\caption{Proposed Deep Learning-based Framework}
	\label{framework_desc}
\end{figure}

\indent This work proposes using deep learning-based models to predict the student performance at the midpoint stage of the course. As is shown in Figure \ref{framework_desc}, the proposed framework includes two main portions.\\ 
The first portion is the data pre-processing and scaling portion. Specifically, the proposed framework uses z-score normalization as per the following equation:
	\begin{equation}
		x_{scaled}=\frac{x-\mu_x}{\sigma_x}
	\end{equation}
where $\mu_x$ is the mean vector and $\sigma_x$ is the standard deviation vector of the training samples. As a result, all features have a unified scale with zero mean and unit standard deviation. It is worth noting that z-score normalization is used in this case since educational data representing grades often follow Gaussian distribution \cite{Nakata}.\\
\indent After splitting the data into training and testing datasets using 80\%-20\% ratio, the data is fed to the deep learning model. Two different models are considered, namely the CNN and the RNN-LSTM models. The CNN model consists of a 1D convolutional layer, a max pooling layer, a flatten layer, and finally a dense layer made up of 128 units with the adam optimizer. The RNN-LSTM model consists of the LSTM layer made up of 64 units and a dense layer made up of 3 units with the ReLu and Softmax activation functions.
\section{Datasets' Description}\label{data_desc}
\indent To evaluate the performance of the proposed deep learning framework, three distinct datasets collected from three different universities located in three separate regions are used. In what follows, a brief description and visualization of these datasets is provided.
\subsection{Dataset 1: First Year Engineering Course at a European University}
\begin{figure}[!ht]
	\centering
	\includegraphics[scale=.4]{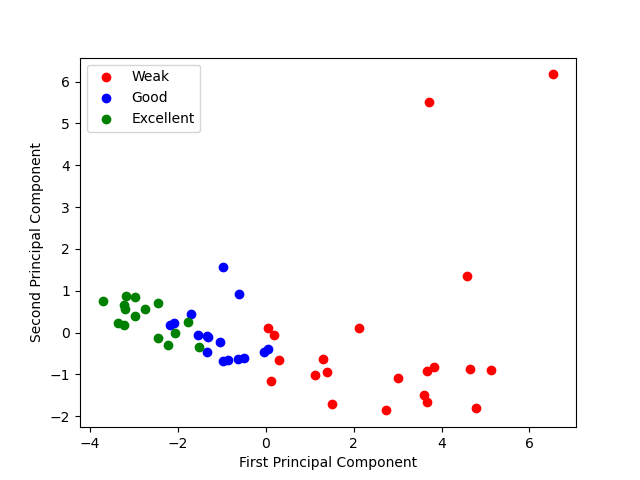}
	\caption{Dataset 1 visualization}
	\label{pca_D1}
\end{figure}

\indent The first dataset was collected for a first year engineering course at a European University \cite{vahdat2015learning}. It consists of a set of tasks of different weights and difficulty levels that were completed using a dedicated simulation environment. Note that the data was collected for 115 students originally. However, only 52 students completed the course and thus were used as part of this dataset. Figure \ref{pca_D1} plots the first and second principal components for Dataset 1.
\subsection{Dataset 2: Third Year Engineering Course at a North American University}
\begin{figure}[!ht]
	\centering
	\includegraphics[scale=.4]{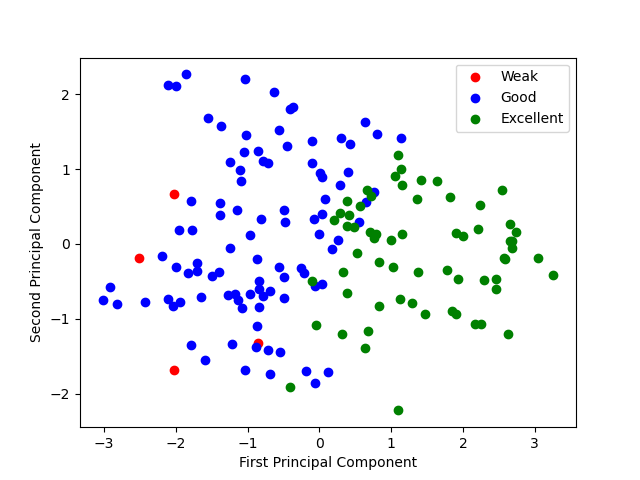}
	\caption{Dataset 2 visualization}
	\label{pca_D2}
\end{figure}

\indent The second dataset was collected for a third year engineering course at a North American university which followed an online delivery mode. It contains the records for 162 students and consists of their performance in various evaluation tasks including assignments, a midterm exam, and a final exam. Figure \ref{pca_D2} plots the first and second principal components for Dataset 2.
\subsection{Dataset 3: First Year IT Course at a Middle Eastern University}
\begin{figure}[!ht]
	\centering
	\includegraphics[scale=.4]{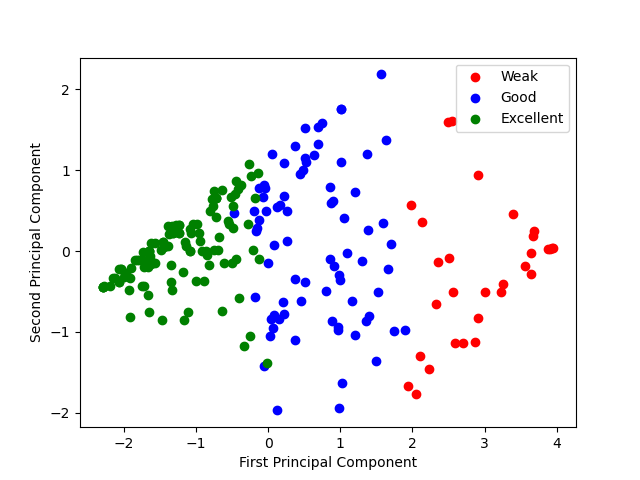}
	\caption{Dataset 3 visualization}
	\label{pca_D3}
\end{figure}

\indent The third dataset was collected for a first year IT course at a Middle Eastern University that was also delivered in an online manner. It contains the records for a group of 222 students and consists of the students’ performance in the different assignments, one midterm exam, and one final exam. Figure \ref{pca_D3} plots the first and second principal components for Dataset 3.
\subsection{Datasets' Processing:}
For all three datasets, the data was further processed by replacing empty marks with a 0 and rounding all decimal value grades to the nearest 1. Moreover, the students' final course grade was used to construct the target classification variable as follows:
\begin{itemize}
    \item Good (G): Given to students that finished the course with a final grade in the range between 70-100\%.
    \item Fair (F): Given to students that finished the course with a final grade in the range between 51-69\%.
    \item Weak (W): Given to students that finished the course with a final grade in the range between 0-50\%.
\end{itemize}
Our goal is to accurately identify the students classified as W since these are the ones that are predicted to potentially fail the course.
\section{Performance Evaluation}\label{performace}
\subsection{Experiment Setup}
\indent \textbf{MATLAB 2022a} is used to plot the PCA figures for the different considered datasets. Additionally, Python is used to develop and evaluate the performance of the proposed framework. %This is done using a laptop having Intel\textsuperscript{\textregistered} Core\textsuperscript{TM} i7-9750H CPU 6 Cores at 2.6 GHZ and 16GB of memory running Windows 10.  
%A Python 3.7.4 running on Anaconda’s Jupyter Notebook is used to evaluate the performance of the proposed framework. The notebook is run on a laptop having Intel\textsuperscript{\textregistered} Core\textsuperscript{TM} i7-9750H CPU 6 Cores at 2.6 GHZ and 16GB of memory running Windows 10.
\subsection{Performance Metrics}
\indent Four main metrics are used to evaluate the performance of the proposed deep learning-based framework. These are the prediction accuracy, precision, recall, and F-score which can be calculated using the following equations \cite{performance_metrics,Injadat_IDS1}:
\begin{equation}
	Accuracy = \frac{(TP + TN)}{(TP + TN + FP + FN)}
\end{equation}
\begin{equation}
	Precision = \frac{TP}{(TP + FP)}
\end{equation}
\begin{equation}
	Recall = \frac{TP}{(TP + FN)}
\end{equation}
\begin{equation}
	F-score = 2\times\frac{(Precision\times Recall)}{(Precision + Recall)}
\end{equation}
\subsection{Performance Results \& Discussion}
\indent The performance of the proposed deep learning-based models is compared to four optimized traditional machine learning (ML) models, namely Support Vector Machines (SVM), K-nearest neighbors (K-NN), Naive Bayes (NB), and Random Forest (RF). These models are optimized using grid search algorithm.\\
\indent Tables \ref{labeled_dataset1_results}, \ref{labeled_dataset2_results}, and \ref{labeled_dataset3_results} provide the performance results of the different ML and DL models for Dataset 1, Dataset 2, and Dataset 3 respectively. The following observations can be made. The results from Table \ref{labeled_dataset1_results} show that the RNN-LSTM model clearly outperformed the other models. In a similar fashion, the results from Table \ref{labeled_dataset3_results} show that the CNN model outperforms the other models. Additionally, although the optimized K-NN model had the best performance for Dataset 2 as is shown in Table \ref{labeled_dataset2_results}, the CNN model achieved comparable performance. The slight degradation of the deep learning-based models for this dataset compared to that of the optimized K-NN is attributed to the extremely low number of weak students to be classified (4 students out of 162) which resulted in these models incorrectly classifying them as good students. 
\begin{table}[!ht]
	\centering
	\caption{Performance Evaluation of Traditional ML and DL Models For Dataset 1}
	\scalebox{0.75}{
		\begin{tabular}{|p{2.2cm}|p{1.3cm}|p{1.3cm}|p{1.3cm}|p{1.1cm}|}	\hline
			Algorithm & Accuracy & Precision & Recall &F-score\\ \hline
			    Optimized SVM&0.45&0.46&0.48&0.46\\ \hline
			Optimized K-NN&0.72&0.59&0.72&0.63\\ \hline
			Optimized RF&0.63&0.56&0.63&0.59\\ \hline
			Optimized NB&0.64&0.71&0.64&0.64\\ \hline
			CNN&0.45&0.56&0.45&0.47\\ \hline
                \textbf{RNN-LSTM}&\textbf{0.82}&\textbf{0.89}&\textbf{0.82}&\textbf{0.82}\\ \hline
		\end{tabular}
	}
	\label{labeled_dataset1_results}
\end{table}

\begin{table}[!ht]
	\centering
	\caption{Performance Evaluation of Traditional ML and DL Models For Dataset 2}
	\scalebox{0.75}{
		\begin{tabular}{|p{2.2cm}|p{1.3cm}|p{1.3cm}|p{1.3cm}|p{1.1cm}|}	\hline
			Algorithm & Accuracy & Precision & Recall &F-score\\ \hline
			    Optimized SVM&0.87&0.88&0.88&0.87\\ \hline
			\textbf{Optimized K-NN}&\textbf{0.94}&\textbf{0.95}&\textbf{0.94}&\textbf{0.94}\\ \hline
			Optimized RF&0.91&0.91&0.91&0.91\\ \hline
			Optimized NB&0.91&0.91&0.92&0.91\\ \hline
			CNN&0.91&0.91&0.91&0.91\\ \hline
                RNN-LSTM&0.88&0.88&0.88&0.88\\ \hline
		\end{tabular}
	}
	\label{labeled_dataset2_results}
\end{table}
\begin{table}[!ht]
	\centering
	\caption{Performance Evaluation of Traditional ML and DL Models For Dataset 3}
	\scalebox{0.75}{
		\begin{tabular}{|p{2.2cm}|p{1.3cm}|p{1.3cm}|p{1.3cm}|p{1.1cm}|}	\hline
		Algorithm & Accuracy & Precision & Recall &F-score\\ \hline
		Optimized SVM&0.88&0.91&0.85&0.88\\ \hline
		Optimized K-NN&0.86&0.87&0.86&0.86\\ \hline
		Optimized RF&0.82&0.82&0.82&0.82\\ \hline
		Optimized NB&0.87&0.86&0.87&0.86\\ \hline
		\textbf{CNN}&\textbf{0.91}&\textbf{0.92}&\textbf{0.91}&\textbf{0.91}\\ \hline
        RNN-LSTM&0.82&0.83&0.82&0.82\\ \hline
		\end{tabular}
	}
	\label{labeled_dataset3_results}
\end{table}

Accordingly, based on the aforementioned results, it can be concluded that deep learning models have significant promise in accurately predicting student performance at the midpoint stage of the online course. Therefore, they can be relied on by university instructors to identify students who are at risk of failing at an earlier stage. In turn, this would result in better student performance as these students can then be guided to offer them further help and ensure they successfully complete and pass the course at hand. 
\section{Conclusion}\label{conc}
\indent Analyzing and evaluating students’ progress in any learning environment is stressful and time consuming if done using traditional analysis methods. With the shift towards integrating the Internet technologies in education and the focus of academic institutions on moving toward e-Learning, blended, or online learning models, this issue is further exasperated. Consequently, the topic of student performance prediction has become a vibrant research area in recent years. To address this, machine learning and data mining techniques have emerged as a viable solution. To that end, this work proposed the use of deep learning techniques, namely convolutional neural networks (CNN) and recurrent neural networks with long short term memory (RNN-LSTM), to predict the students' performance at the midpoint stage of the online course delivery using three distinct datasets collected from three different regions of the world. Experimental results showed that deep learning models had promising performance. More specifically, they outperform other optimized traditional ML models in two of the three considered datasets while also having comparable performance for the third dataset.\\
\indent In terms of model performance, the main takeaway is that no single deep learning model can be generalized across multiple datasets. Therefore, it is recommended that multiple models are combined as ensembles using different techniques (e.g. majority voting, bagging, boosting, etc.) to ensure that the student performance prediction is performed more accurately and in a more generic manner.\\
\indent From the application point of view, these promising results open the door to offering more customized and personalized learning experiences. More specifically, the results achieved using the aforementioned framework open the door to further analysis. For example, the data associated with students that are predicted to fail the course can be further studied and analyzed to determine the root cause of their poor performance. This can be done using techniques such as association rules to determine which learning outcomes are these students struggling with. Subsequently, additional resources can be provided to these students (e.g. additional videos, additional readings, practice exercises, etc.) to ensure that they are meeting the expected student learning outcomes, making their learning experience feel better customized and personalized.

\small
\balance
\bibliographystyle{IEEEtran}
\bibliography{sample}

\end{document}